\documentclass[11pt,twoside]{article}
\usepackage{./asp2014}

\aspSuppressVolSlug
\resetcounters

\begin{document}

\title{Untangling the White Dwarf Luminosity Functions}
\author{M. C. Lam
\affil{Institute for Astronomy, University of Edinburgh, Edinburgh, UK \email{mlam@roe.ac.uk}}
}

\paperauthor{M. C. Lam}{mlam@roe.ac.uk}{0000-0002-9347-2298}{Institute for Astronomy}{University of Edinburgh}{Edinburgh}{Midlothian}{EH9 3HJ}{UK}

\begin{abstract}
The inversion of the white dwarf luminosity function provides an independent way to prove the past star formation history of the Milky Way independent of any cosmological models. In Rowell \& Hambly~(2011), the effective volume method uses the average properties of all the objects in a given bin, so a significant amount of information is lost in the early stage of the analysis, in this work, I explore the possibility of assigning objects individually in a probabilistic way using the generalised Schmidt density estimator~($1/\mathrm{V}_{\mathrm{max}}$).
\end{abstract}

\section{Motivation}
The white dwarf luminosity function of the solar neighbourhood has been studied for decades. However, due to the lack of chemical signature from the progenitor, these degenerate objects from all three components of the Milky Way in the solar neighbourhood have been studied all together. Works have been assuming thin disc properties because they completely dominate the sample; or separating the thin and thick disc sample from the halo objects by applying large tangential velocity limits. In this work, I will demonstrate the potential of separating the three components probabilistically to study the individual luminosity functions. The untangling of the three populations is essential to the detailed study of star formation history through the inversion of each luminosity function~(Rowell~2013).

\section{Population Synthesis}
Monte Carlo simulations are used to produce snapshots of the WD-only solar neighbourhoods which carry six-dimensional phase-space information of the thin disk~(8\,Gyr old with constant star formation), thick disk~(10\,Gyr old with constant star formation) and stellar halo~(12.5\,Gyr old with 1\,Gyr of constant star formation in the beginning). For details of the model, please refer to Section 2 of Lam, Rowell \& Hambly~(2015). There is only one change applied to the model: the thick disc scaleheight is set at 1000\,pc~(Sandage \& Fouts 1987) instead of 750\,pc.

\section{White Dwarf Luminosity Function}
The generalised Schmidt Estimator~($1/\mathrm{V}_{\mathrm{max}}$) is used to construct the white dwarf luminosity function. With this method, the contribution of each object into the luminosity function is weighted by the inverse of the maximum volume in which the object could be observed by the survey. However, the original treatment assumes objects are uniformly distributed. In reality, stars in the solar neighbourhood are concentrated in the plane of the disk so the effects of space-density gradient have to be corrected. This led to the development of the generalised volume $\mathrm{V}_{\mathrm{gen}}$~(Stobie, Ishida \& Peacock 1989; Tinney, Reid \& Mould 1993) which is calculated by integrating the appropriate stellar density profile $\rho/\rho_{\odot}$ along the line of sight, and further corrected for kinematics limits by Lam, Rowell \& Hambly (2015),
\begin{equation}
\mathrm{V}_{\mathrm{gen}} = \Omega \int^{\mathrm{D}_{\mathrm{max}}}_{\mathrm{D}_{\mathrm{min}}} r^{2} \times \frac{\rho(r)}{\rho_{\odot}} \times \left[ \int^{b(r)}_{a(r)} P(v_{\mathrm{tan}}) \mathrm{d}v_{\mathrm{tan}} \right] \mathrm{d}r
\end{equation}
where $\Omega$ is the size of the solid angle of the survey, the limits of the integral are $\mathrm{D}_{\mathrm{min}}$ and $\mathrm{D}_{\mathrm{max}}$ which are the distances limits set by the bright and faint photometric limits, the inner integral is to correct for incompleteness due to kinematic selection by integrating over the allowed range of tangential velocities restricted by the distance limits. The number density at a given magnitude is given by
\begin{equation}
\phi(\mathrm{mag}) = \sum_{i} \frac{1}{\mathrm{V}_{\mathrm{gen},i}}
\end{equation}
and the associated variance is
\begin{equation}
\sigma_{\phi}(\mathrm{mag}) = \sum_{i} \left( \frac{1}{\mathrm{V}_{\mathrm{gen},i}} \right)^{2}.
\end{equation}

\section{Membership Likelihood}
The little information WDs reveal from their atmospheres makes the task of separating them into different populations extremely difficult. The most notable differences between the three populations are the kinematic structures. Given that most objects do not have radial velocity measurements~(and for the cool WDs, their spectra are smooth), one can only rely on the projected velocities. Thus, the tangential velocity is used to associate objects to the host population. In a simple but sufficient picture of the solar neighbourhood, the tangential velocity can be approximated by marginalising the Schwarzschild distribution onto the plane of observation~(Rowell \& Hambly 2011). This distribution is a function of position so it has to be calculated for each small part of the sky. By taking into account the observability of the survey, the effective tangential velocity distribution function, $P'(v_{\mathrm{tan}})$, can be constructed for each of the three populations for different parts of the sky. The probability of each object being drawn with each population is thus given by
\begin{equation}
P_{\mathrm{pop}} = \frac{1}{\sqrt{2\pi\sigma^{2}_{v_{\mathrm{tan}}}}} \int^{\infty}_{0} n_{\mathrm{pop}} \times P'(v_{\mathrm{tan}}) \times \mathrm{exp}\left[ \frac{(v'_{\mathrm{tan}}-v_{\mathrm{tan}})^{2}}{\sigma^{2}_{v_{\mathrm{tan}}}} \right] \mathrm{d}v'_{\mathrm{tan}} 
\end{equation}
where $n_{\mathrm{pop}}$ is the number density of the population, the distribution of the uncertainty in the tangential velocity is assumed to be Gaussian. The final probabilities of each object are 
\begin{align}
P_{\mathrm{thin}} &= P_{\mathrm{thin}} / ( P_{\mathrm{thin}} + P_{\mathrm{thick}} + P_{\mathrm{halo}} ), \\
P_{\mathrm{thick}} &= P_{\mathrm{thick}} / ( P_{\mathrm{thin}} + P_{\mathrm{thick}} + P_{\mathrm{halo}} ) \textrm{ and}\\
P_{\mathrm{halo}} &= P_{\mathrm{halo}} / ( P_{\mathrm{thin}} + P_{\mathrm{thick}} + P_{\mathrm{halo}} ).
\end{align}
Since the number densities of the components are not known, they have to be set as free parameters together with the shape of the WDLF~(i.e.\ number density as a function of magnitude). The input density, $n_{\mathrm{pop},i}$ of the component densities adopt $90\%$, $10\%$ and $2\%$ that of the thin disk assuming all objects belong to the thin disk, and in each iteration, $n_{\mathrm{pop},i}$ is updated. They do not sum to $100\%$ because these are only crude estimate. Furthermore, when the three components are untangled, it is expected that the total normalisation will go down because the scaleheight correction of the thin disc will over estimate the densities when thick disc and halo objects are corrected with the thin disc geometry. The measured number density is modified to
\begin{equation}
\phi_{\mathrm{pop},m} = \sum_{i} \frac{P_{\mathrm{pop}}}{\mathrm{V}_{\mathrm{gen},i}}
\end{equation}
and the associated variance is
\begin{equation}
\sigma_{\phi_{\mathrm{pop},m}}^{2} = \sum_{i} \left( \frac{P_{\mathrm{pop}}}{\mathrm{V}_{\mathrm{gen},i}} \right)^{2}.
\end{equation}
The $\chi^{2}$ expression to be minimised is
\begin{equation}
\chi^{2} = \left( \frac{ \phi_{\mathrm{thin},m} - \phi_{\mathrm{thin},i} }{\sigma_{\phi_{\mathrm{thin},m}}} \right)^{2} + \left( \frac{ \phi_{\mathrm{thick},m} - \phi_{\mathrm{thick},i} }{\sigma_{\phi_{\mathrm{thick},m}}} \right)^{2} + \left( \frac{ \phi_{\mathrm{halo},m} - \phi_{\mathrm{halo},i} }{\sigma_{\phi_{\mathrm{halo},m}}} \right)^{2}.
\end{equation}

\section{Preliminary Results}
The thin and thick disc luminosity functions are at roughly the correct normalisation~(see Figure 1). However, the halo density is overestimated by an order of magnitude. This could be due to the strong contrast in densities between the thin disc and halo, causing significant contaminations to each other: thin disk objects have too high probabilities to be halo objects so the halo luminosity function is much enhanced, while the probability of the relatively slow halo objects being drawn from the thin disc is too high, and the density correction for thin disc further amplifies the contamination.
\articlefigure{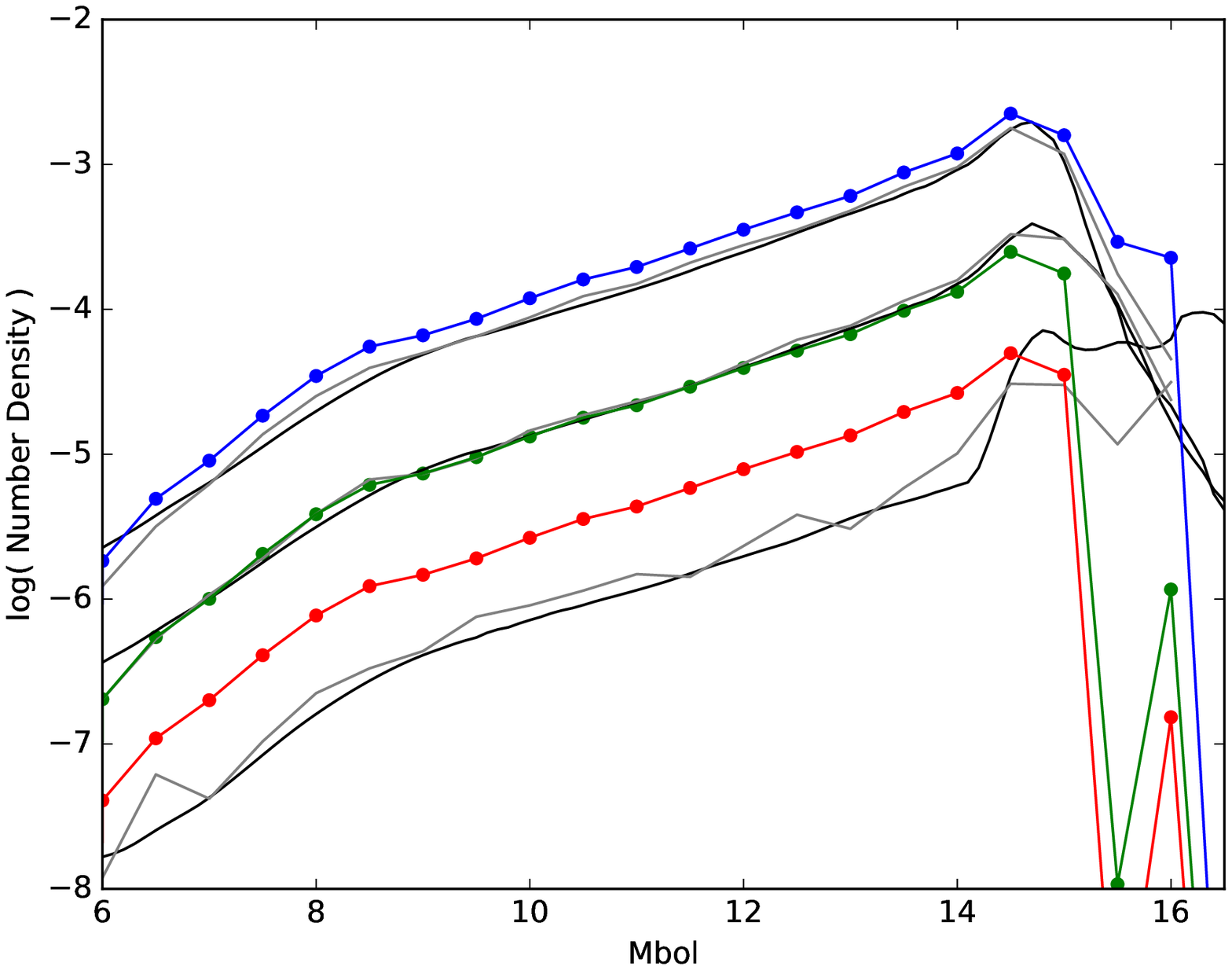}{}{The input theoretical luminosity functions are in black. Individual luminosity functions if all objects are perfectly classified are in grey, and the recovered functions are indicated by the lines overplotted with scattered points~(top - thin disc, middle - thick disc, bottom - halo).}

\section{Future Work}
Simulations with different number density and star formation histories should be tested for the stability of this method. More parameters may be introduced to help separate the halo objects from the other two components. The uncertainties in the number density as a function of magnitude have to be calculated explicitly. If this method can successfully untangle the components, the technique should be transferred to the maximum likelihood density estimators such that the analysis can be done in a seamless process.

\acknowledgments{ML acknowledges financial support from the IfA~(Edinburgh) Consolidated Grant and useful discussion with N.~C.~Hambly and N.~Rowell.}


\end{document}